\documentclass[prd,twocolumn,showpacs,showkeys,preprintnumbers,floatfix,
nofootinbib,superscriptaddress]{revtex4}
\usepackage{amsfonts} 
\usepackage{amssymb} 
\usepackage{amsmath} 
\usepackage{graphicx} 
\usepackage{subfigure} 
\usepackage{array} 
\usepackage{dcolumn} 
\usepackage{bm} 
\usepackage{latexsym} 
\usepackage{longtable} 
\usepackage{hyperref} 
\graphicspath{{./figs/}}
\renewcommand{\tilde}{\widetilde}
\newcommand{\tr}{\textrm{tr}}
\newcommand{\Tr}{\textrm{Tr}}
\newcommand{\Str}{\textrm{Str}}

\begin{document}


\title{Constraint on the Low Energy Constants
of Wilson Chiral Perturbation Theory}

\author{Maxwell T. Hansen}
\email[Email: ]{mth28@uw.edu}
\affiliation{
 Physics Department, University of Washington, 
 Seattle, WA 98195-1560, USA \\
}
\author{Stephen R. Sharpe}
\email[Email: ]{sharpe@phys.washington.edu}
%
%
\affiliation{
 Physics Department, University of Washington, 
 Seattle, WA 98195-1560, USA \\
}
%
%
%
%
%
%
\date{\today}
\begin{abstract}
Wilson chiral perturbation theory (WChPT)
is the effective field theory describing the
long-distance properties of lattice QCD with 
Wilson or twisted-mass fermions.
We consider here WChPT for the theory with
two light flavors of Wilson fermions or a single light twisted-mass fermion.
Discretization errors introduce three low energy constants (LECs)
into partially quenched WChPT at ${\cal O}(a^2)$, conventionally
called $W_6'$, $W_7'$ and $W_8'$. The phase structure of the
theory at non-zero $a$ depends on the sign of the combination
$2 W'_6+W'_8$, while the spectrum of the lattice Hermitian
Wilson-Dirac operator depends on all three constants.
It has been argued, based on the positivity of 
partition functions of fixed topological charge, 
and on the convergence of graded group integrals
that arise in the $\epsilon$-regime of ChPT,
that there is a constraint on the
LECs arising from the underlying lattice theory. In particular,
for $W_6'=W_7'=0$, the constraint found is $W'_8\le0$.
Here we provide an alternative line of argument, based on mass inequalities
for the underlying partially quenched theory.
We find that $W'_8\le0$, irrespective of the values of $W'_6$
and $W'_7$. 
Our constraint implies that $2 W'_6> |W'_8|$
if the phase diagram is to be described by the first-order
scenario, as recent simulations suggest is the case for some
choices of action.

\end{abstract}
\pacs{12.38.Gc, 11.30.Rd, 12.15.-y}
\keywords{lattice QCD, pion scattering, discretization errors}
\maketitle

\section{Introduction \label{sec:intr}}

Effective field theories such as chiral perturbation theory
(ChPT) contain coefficients, usually called
low energy constants (LECs), which are not determined by symmetry.
If the matching between the high and low-energy theories is
non-perturbative, as is the case in the matching of QCD to ChPT, 
then the LECs must be determined either by experiment or
by a non-perturbative method such as lattice QCD.
One usually has no information on the LECs,
other than a prediction for
their order of magnitude based on naive dimensional analysis.
It is sometimes possible, however, to constrain the signs of 
particular LECs based on the physics of the high-energy theory.
For example, certain four-derivative terms in the chiral Lagrangian
are constrained to be positive based on causality~\cite{PhamTruong}.
This argument has been generalized and applied widely in Ref.~\cite{A_Adams}.
Another example concerns the chiral Lagrangian describing a lattice
simulation at non-zero lattice spacing with a mixed-action 
(different valence and sea-quark actions).
It is found in Ref.~\cite{BarGS} that, using 
generalized QCD mass inequalities~\cite{mass-inequalities},
one finds a constraint on a combination of the
LECs which arise due to discretization errors.

A further method of constraining LECs has recently been
discovered in the context of calculating the low-energy
spectrum and eigenvalue properties of the lattice
Hermitian Wilson-Dirac operator~\cite{DSV,ADSV}.
One line of argument is based on the positivity of 
the underlying two-flavor fermion determinant, which follows from the
$\gamma_5-$hermiticity of the Wilson-Dirac operator.
Specifically, the partition function at
fixed (odd) topology is positive in the underlying
theory but is only positive in the effective theory
(here partially quenched Wilson ChPT [PQWChPT])
if the LECs satisfy a constraint~\cite{ADSV}. 
In the standard convention for LECs,\footnote{%
Note that our convention for $W'_j$, which follows
Ref.~\cite{BarRS03}, differs in sign from that used in
Refs.~\cite{DSV,ADSV}.}
this constraint is $W'_8\le0$ if $W'_6=W'_7=0$.\footnote{%
Since these LECs appear at leading order in the appropriate
power-counting, they are independent of the renormalization
scale.}
Another line of argument notes that the partially-quenched
partition function for zero-momentum modes
(which determines the leading order behavior in the $\epsilon$-regime)
converges only if $W'_8\le W'_6+W'_7$~\cite{DSV,ADSV}.\footnote{%
It may be possible to obtain
further constraints from these or similar lines of 
argument~\cite{Jac_ECT,ADSV}.}
We also note that a similar constraint
(specifically, $W'_8\le 0$ independent of $W'_6$ and $W'_7$)
was found earlier by one of us, based on the finding
that the method for calculating the spectral density
in infinite volume using PQWChPT only worked if the
constraint held~\cite{gap06}. It was not clear, however,
whether this was a fundamental constraint or simply
a shortcoming of the method of calculation.

The constraints found in Refs.~\cite{DSV,ADSV,gap06} 
imply an interesting corollary if one assumes that scaling 
at large $N_c$ (number of colors) is a good guide at $N_c=3$. 
In particular, since $W'_6/W'_8 \sim W'_7/W'_8 \sim 1/N_c$,
this assumption would mean that one can ignore $W'_6$ and
$W'_7$ to first approximation.
Then the constraints imply that {\em any}
discretization of Wilson fermions will have an Aoki phase
for small enough physical quark mass.
The other possible phase 
diagram---the first-order scenario~\cite{SS}---would not occur.
This is in apparent contradiction with the results of
simulations using dynamical twisted-mass fermions, 
which find strong evidence for the first-order
scenario~\cite{Farchioni:2004us,Farchioni:2004ma,Farchioni:2004fs,%
Farchioni05,Farchioni:2005bh,Farchioni06,Boucaud:2007uk,ETM08,ETM09,%
Baron:2010bv}.
Of course, large $N_c$ scaling may not be useful for $N_c=3$, 
in which case the
connection between the constraint and the phase scenario
need not hold. 

Given this situation, we think that it is important
to find an alternative 
line of argument leading to such constraints.
This is what we provide in the present note.
In particular, we find that
a generalization of the mass-inequality method
of Ref.~\cite{BarGS} constrains the LECs of WChPT.\footnote{%
The fact that mass-inequalities can provide useful
information in twisted-mass theories has also been noted in
Refs.~\cite{Jac_ECT,Jac_Kim_inprep}.}
Our constraint results from considering the
twisted-mass generalization of Wilson fermions
and comparing the {\em quark-connected} part of the
neutral pion propagator to the charged pion propagator.
A partially quenched set-up
is required to separate the quark-connected and disconnected
contractions, and this is why it is the LECs of 
{\em partially quenched} WChPT which enter.
We find $W'_8\le0$, independent of $W'_6$ and $W'_7$.

The remainder of this note is organized as follows.
In the following section we explain how partial quenching
allows one to separately calculate the quark-connected
part of the neutral pion correlator.
In Sec.~\ref{sec:WChPT} we present the
calculation of the quark-connected neutral pion ``mass''
at leading order in WChPT. We do so only at maximal twist,
since this suffices to show the constraint.
In Sec.~\ref{sec:inequalities} we derive an
inequality among quark-connected correlation
functions, from which follows the above-noted
constraint. We summarize and offer some 
concluding comments in Sec.~\ref{sec:conc}.
We relegate some technical details to two appendices,
the first concerning the form of the condensate in the
partially quenched theory, and the second extending
the analysis of the main text from maximal 
to arbitrary twist.

\section{Using partial quenching to select quark-connected correlators}
\label{sec:PQ}

Our argument uses twisted-mass fermions~\cite{Frezzotti99,Frezzotti03},
so we begin by recalling the salient features of this approach.
In an unquenched theory, the quark Lagrangian takes the form
\begin{equation}
{\cal L}_q =
\bar q_S (D_W + m_0 + i\mu_0 \gamma_5\tau_3) q_S
\,,
\label{eq:Lq}
\end{equation}
where $q_S$ is an isodoublet of quark fields (corresponding
to the up and down quarks), 
and $D_W$ is the Wilson-Dirac operator.
The subscript ``S'' indicates that these are
sea-quarks, appearing in the fermion determinant, as opposed
to the valence quarks introduced below.\footnote{%
Simulations using two doublets of dynamical twisted-mass fermions are
also now being done, with the second such fermion describing
the strange and charm quarks~\cite{Baron:2010bv}. 
The arguments in this note apply equally well to such a set-up,
however, because the second doublet contains degrees of freedom
that are heavy on the scale of the light up and down quarks.
Thus the form of the chiral Lagrangian used in Sec.~\ref{sec:WChPT}
is unchanged (although the values of the LECs will be different),
and the argument for the mass inequalities in Sec.~\ref{sec:inequalities}
goes through unchanged.
It is important in this regard that the determinant in such
$N_f=2+1+1$ simulations remains real and positive~\cite{Frezzotti:2003xj}.}
We refer to $m_0$ as the normal (bare) mass 
and $\mu_0$ as the twisted (bare) mass.
The following considerations do not depend on whether
$D_W$ is improved, or on the form of the gauge action,
so we do not specify either. We will need
only the property of ``$\gamma_5$-hermiticity'':
\begin{equation}
\gamma_5 D_W \gamma_5 = D_W^\dagger
\,.
\end{equation}
When writing the Lagrangian in the form (\ref{eq:Lq}), we
are using what is commonly called the ``twisted basis'',
in which the mass, and not the Wilson term, is twisted.

In the continuum limit, a mass term $m+i\mu\gamma_5\tau_3$
can be rotated into a purely normal mass $m_{q}=\sqrt{m^2+\mu^2}$
by an appropriate axial rotation. Thus the apparent breaking
of flavor by the $\mu$ term is misleading---%
flavor is preserved for all $\mu$.
At non-zero lattice spacing, however, flavor is 
explicitly broken
from $SU(2)$ to $U(1)$, leading to a splitting of the pion
multiplet: $m_{\pi^\pm}\ne m_{\pi^0}$. As is well known
(and as will be seen explicitly in the following section)
the splitting is of ${\cal O}(a^2)$.

The particular case of maximal twist corresponds to
tuning $m_0\to m_c$ such that the {\em physical} normal mass
vanishes (or is, at least, sufficiently small compared to
the twisted mass). There are a number of different
tuning criteria that can be used, leading to results
for physical quantities differing only at ${\cal O}(a^2)$.
For discussion of these issues see
Refs.~\cite{Jansen:2003ir,Aoki:2004ta,Frezzotti:2005gi,Jansen:2005gf,%
Jansen:2005kk,Aoki:2006nv,SharpeNara06,Boucaud:2007uk,ETM08,Baron:2010bv}.
All that matters here, however, is that a consistent
criterion exists in which $m_c$ is fixed,
such as the one based on the PCAC mass used in practice in 
present simulations~\cite{ETM08,ETM09,Baron:2010bv}.

We will also need to know the quark-level operators which
couple to the charged and neutral pions in the twisted basis.
These are given, e.g., in Appendix A of Ref.~\cite{ETM08}.
The charged pions are created by
\begin{equation}
P^{\pm} = i \bar q_S \gamma_5 \tau_{\pm} q_S
\qquad
[\tau_{\pm} = \frac1{\sqrt{2}} (\tau_1 \pm i \tau_2)]
\label{eq:Ppm}
\end{equation}
(independent of twist angle),
while the neutral pion is created at maximal twist by
\begin{equation}
S^0 = - \bar q_S q_S
\,.
\label{eq:S0}
\end{equation}
Thus the two-point correlators of the charged fields,
\begin{equation}
C^\pm(n) = \langle P^\mp(0) P^\pm(n) \rangle
\label{eq:Cpm}
\end{equation}
($n$ labeling lattice sites),
have only quark-connected contributions.
For example, 
\begin{equation}
C^+(n) = 2\langle {\rm tr}\left(\gamma_5
G(\mu)_{0,n} \gamma_5 G(-\mu)_{n,0} \right) \rangle
\,,
\label{eq:C+initial}
\end{equation}
where the trace is over (implicit) color and Dirac
indices, and the quark propagator is
\begin{equation}
G(\mu)_{0,n} = \left(\frac{1}{D_W + m_c + i \mu\gamma_5}\right)_{0,n}
\,.
\end{equation}
The neutral pion propagator, however, has both quark-connected
and disconnected contributions
\begin{eqnarray}
C^0(n) &=& \langle S^0(0) S^0(n) \rangle
\label{eq:C0}
\\
&=& C^{0,{\rm conn}}(n) + C^{0,{\rm disc}}(n)
\\
C^{0,{\rm conn}}(n) &=& -\Big\langle
{\rm tr}\big(G(\mu)_{0,n} G(\mu)_{n,0}
\nonumber\\
&&\qquad\qquad +G(-\mu)_{0,n}G(-\mu)_{n,0}\big)
\Big\rangle
\label{eq:C0conn_initial}
\\
C^{0,{\rm disc}}(n) &=& \Big\langle
{\rm tr}\left(G(\mu)_{0,0}+G(-\mu)_{0,0}\right)
\nonumber\\
&&\qquad \times 
{\rm tr}\left(G(\mu)_{n,n}+G(-\mu)_{n,n}\right) \Big\rangle
\,.
\end{eqnarray}
In practice, the quark-connected part has a much better signal to noise
ratio than the disconnected part, but improved techniques have allowed
the computation of the latter with errors small enough that the
mass of the neutral pion can be extracted~\cite{ETM08}.

What we are interested in here, however, is the quark-connected
part of the correlator. Since this is measured with small errors
(comparable to those for the charged correlator)
it is worthwhile investigating whether it contains
useful information.
In the physical two-flavor theory one cannot separate the
two Wick contractions. It is well known, however, that if one
considers the partially quenched (PQ) extension of the theory~\cite{BGPQ}, 
then, by adding enough valence quarks, one can find correlation
functions which pick out any desired Wick contraction.
In the present case it suffices to add a single valence
isodoublet $q_V$ and the corresponding ghost quark isodoublet
$\tilde q_V$. The Lagrangian for each of these quarks is the
same as that for $q_S$ [the twisted-mass Lagrangian (\ref{eq:Lq})],
except that $\bar q_S$ is replaced by $\tilde q_S^\dagger$
for the ghost quark.\footnote{%
This glosses over an important subtlety. In the ghost sector,
convergence of the functional integral requires that the real
part of the eigenvalues of the discretized fermion operator
are positive. This is not the case for
$D_W + m_0 + i\mu\tau_3\gamma_5$ given that one always works
with $m_0<0$. This issue has been resolved, in the context of
the quenched theory, in Ref.~\cite{GSS}, and a simple generalization
works here. The solution is to do an axial rotation in the $\tau_3$
direction by angle $\pi/4$, such that the fermion operator becomes
$D -i \gamma_5\tau_3(W+m_0) + \mu$, where $D$ is the naive
discretization of the Dirac operator and $W$ the Wilson term.
This new operator consists of an antihermitian part, with purely
imaginary eigenvalues, and a real offset $\mu$, which we choose to be
positive. (For negative $\mu$, an axial rotation in the other direction
resolves the problem.) For maximal twist, this is exactly the axial
rotation that brings one to the physical basis, but for other twist
angles it gives a different basis. In this new basis one can add in
valence and ghost quarks. One then goes through the standard steps
to obtain the chiral Lagrangian including discretization errors~\cite{SS},
following a simple generalization of the analysis of Ref.~\cite{GSS}.
Compared to the usual chiral Lagrangian, one has additional factors
of $\pm i\tau_3$ in terms containing spurions coming from discretization
errors. Thus the Lagrangian looks non-standard.
In the quark sector (sea and valence) one can, however, undo the
axial rotation (now at the level of the chiral fields)
ending up with the standard form of the chiral Lagrangian for WChPT
[presented below in Eq.~(\ref{eq:L})]. This does not work in the
ghost sector, since one is not allowed to do normal axial
rotations there. This restriction does not, however, 
effect the present calculation,
since we only consider correlation functions in the quark sector.
In fact, the correct procedure in the ghost sector has been
worked out in Ref.~\cite{Necco}, generalizing the methodology
of Ref.~\cite{GSS}.}

Within this PQ setup, the correlation function which yields the
quark-connected part of the neutral pion correlator involves the
mixed valence-sea pion:\footnote{%
We could just as well add two isodoublets of valence quarks
(and corresponding ghosts) and use
$\langle \bar q_{V1} q_{V2}(0)\bar q_{V2}q_{V1}(n)\rangle$.
The choice made in the text is, however, the minimal one.}
\begin{equation}
C^{0,{\rm conn}}(n) =
\langle \bar q_S q_V(0)\ \bar q_V q_S(n) \rangle
\,.
\label{eq:C0conn}
\end{equation}
This is because there is no disconnected Wick contraction
between a $\bar q_S$ and $q_V$.
Although we will not use it, it is perhaps of interest to
note that the disconnected neutral pion Wick contraction can
be obtained as
\begin{equation}
C^{0,{\rm disc}}(n) =
\langle \bar q_S q_S(0)\ \bar q_V q_V(n) \rangle
\,.
\end{equation}

Partial quenching is often used to consider valence masses
(or actions) differing from those of the sea quarks. 
In this work, by contrast, the valence and sea quarks have
identical actions and masses. Thus there is an exact
$SU(2)$ flavor symmetry mixing valence and sea quarks.
This is a subgroup of the $SU(4)$ flavor symmetry that emerges
in the continuum limit (itself a subgroup of the graded 
flavor group $U(4|2)$ that holds for perturbative calculations
in the continuum PQ theory~\cite{Sharpe:2001fh}).

We also remark that $C^{0,{\rm conn}}$ and $C^{0,{\rm disc}}$ are
separately unphysical---they cannot be expressed in terms of
a sum of exponentially falling terms with positive (real) coefficients.
Nevertheless, they can be calculated using the appropriate
low-energy effective theory---PQWChPT---which itself is an
unphysical theory, although perfectly well defined in Euclidean
space. It turns out that $C^{0,{\rm conn}}$ does have,
at leading order in WChPT, a physical form at long distances,
which is all that we need for our argument.

\section{Wilson ChPT calculation of connected pion masses}
\label{sec:WChPT}

In this section we calculate $C^{0,{\rm conn}}$ and
$C^\pm$ using PQWChPT. We are interested in the
differences between these two correlators,
which turn out to arise at ${\cal O}(a^2)$.
Thus we work to leading order (LO)
in the ``large cut-off effects'' or ``Aoki'' regime in which
the power counting is $m\sim \mu\sim a^2\Lambda_{\rm QCD}^3$,
where $m$ and $\mu$ are the physical normal and twisted masses
[defined in Eq.~(\ref{eq:m_mu_def}) below].
There is no need to work to higher order, since for the
purposes of constraining the LECs we can imagine that $m$,
$\mu$ and $a^2$ are arbitrarily small.

At leading order, and after shifting the quark mass to remove
an ${\cal O}(a)$ term, 
the partially quenched chiral Lagrangian is~\cite{SS,BarRS03}
\begin{eqnarray}
{\cal L}_\chi &=&
\frac{f^2}4 \Str\left(
 \partial_\mu \Sigma \partial_\mu \Sigma^\dagger\right)
-
\frac{f^2}4 \Str\left( \chi^\dagger \Sigma+\Sigma^\dagger \chi\right)
\nonumber\\
&& - \hat{a}^2 W_6' \left[\Str\left(\Sigma+\Sigma^\dagger\right)\right]^2
 - \hat{a}^2 W_7' \left[\Str\left(\Sigma-\Sigma^\dagger\right)\right]^2
\nonumber\\
&&
 - \hat{a}^2 W_8' \Str\left(\Sigma^2+[\Sigma^\dagger]^2\right)
\,.
\label{eq:L}
\end{eqnarray}
Here $\Sigma\in SU(4|2)$, ``Str'' stands for supertrace, 
$\chi = 2 B_0 M$
with $M$ the mass matrix, and $\hat a = 2 W_0 a$. 
$B_0$ and $f$ are continuum LECs 
(with the convention $f_\pi\approx 93\;$MeV), while $W_0$,
$W'_6$, $W'_7$ and $W'_8$ are LECs associated with discretization
errors.
Since our set-up requires just two valence quarks and two ghosts,
the graded chiral symmetry is\footnote{%
The actual symmetry differs
from this due to the constraints from convergence of ghost integrals.
For perturbative calculations, such as those we perform
here, one can, however, work as if the symmetry is as claimed.
This was shown for the continuum PQ theory in Ref.~\cite{Sharpe:2001fh},
and presumably carries over to Wilson PQChPT.
In fact, all we need in the present calculation are fluctuations
in the quark sector, and here the appropriate symmetry is certainly
$SU(4)$. For non-perturbative calculations, however,
such as those done in Refs.~\cite{DSV,ADSV}, one must account
for the need to have convergent integrals in the ghost sector,
which leads to a different global group.}
$SU(4|2)_L\times SU(4|2)_R$.

The mass matrix in ${\cal L}_\chi$ is related to the bare masses
in the underlying quark Lagrangian (\ref{eq:Lq}).
For the unquenched theory, we have
\begin{eqnarray}
M &=& m+i\mu\tau_3 = m_q e^{i\omega_m\tau_3}
\\
m &=& Z_S^{-1}(m_0-m_c)/a\,,\quad
\mu = Z_P^{-1} \mu_0/a\,.
\label{eq:m_mu_def}
\end{eqnarray}
Here $\omega_m$ is the ``input'' twist angle, and
$m_q=\sqrt{m^2+\mu^2}$ is the physical mass in the continuum limit.
Maximal twist corresponds to $m=0$. Note that in this case one
does not need to know the renormalization factors $Z_S$ and $Z_P$
in order to determine the twist angle.
For the partially quenched theory, the mass matrix,
which has dimension $6\times 6$, is block diagonal,
with each block containing $M$.
As noted in the previous section, this mass matrix
leaves an unbroken $SU(2)$ symmetry between sea and valence
quarks.


We must first determine the orientation of the vacuum,
$\Sigma_0=\langle0|\Sigma|0\rangle$, taking into
account the ${\cal O}(a^2)$ terms.
In the unquenched sector this has been done in
Refs.~\cite{Munster,Scorzato,ShWu1}.
Writing
%
%
\begin{equation}
\Sigma_0^{\rm unqu} =e^{i\omega_0\tau_3}
\,,
\end{equation}
one needs in general to solve a quartic 
[given in Eq.~(\ref{eq:nolinear})]
to determine $\omega_0$, and $\omega_0-\omega_m$ is
generically of ${\cal O}(1)$. For the special case of
maximal twist, however,
the solution is simply $\omega_0=\omega_m=\pm \pi/2$,
i.e. the input and output twist angles are the same.

For the partially quenched theory, we argue
in Appendix~\ref{app:vafawitten} that
\begin{equation}
\Sigma_0 = \left(\begin{array}{ccc}
e^{i\omega_0\tau_3} & 0 & 0 \\
0 & e^{i\omega_0\tau_3} & 0 \\
0 & 0 & e^{\phi_g} \end{array}\right)
\,,
\label{eq:Sigma0}
\end{equation}
in a $2\times2$ block notation with the blocks
ordered as sea, valence and ghost quarks.
In words, this result says 
that the $SU(2)$ valence-sea symmetry is unbroken
(one implication of which is that the vacuum twist in the
valence sector is the same as that in the sea sector) 
and that there are no quark-ghost condensates.
We do not need to discuss the (subtle issue) of the ghost
condensate $e^{\phi_g}$ (which is a $2\times2$ matrix), 
since we will not need propagators
involving ghosts. This issue has been discussed, albeit in
a different power-counting, in Ref.~\cite{Necco}.

Pion masses can now be obtained by considering
small oscillations around the condensate.
To do this, we use
\begin{equation}
\Sigma = \xi_0 \Sigma_{\rm ph} \xi_0\,,\quad
\xi_0 = 
\left(\begin{array}{ccc}
e^{i\omega_0\tau_3/2} & 0 & 0 \\
0 & e^{i\omega_0\tau_3/2} & 0 \\
0 & 0 & e^{\phi_g/2} \end{array}\right)
\label{eq:expandsigma}
\end{equation}
with $\Sigma_{\rm ph} = \exp(i\sqrt2 \pi/f)$ containing 
the pion fields.
We will only need the quark part of the pion field, which
we decompose as follows
\begin{equation}
{\cal P}_q \pi {\cal P}_q = 
\left(\begin{array}{ccc}
\pi_{SS} & \pi_{SV} & 0 \\
\pi_{VS} & \pi_{VV} & 0 \\
0 & 0 & 0 \end{array}\right)
\,,
\label{eq:pi_q}
\end{equation}
with ${\cal P}_q$ the projector onto the quark subspace
\begin{equation}
{\cal P}_q =
\left(\begin{array}{ccc}
1 & 0 & 0 \\
0 & 1 & 0 \\
0 & 0 & 0 \end{array}\right)
\,.
\end{equation}
The block pion fields in (\ref{eq:pi_q}) contain isosinglet
components (i.e. $\eta$-like fields) as well as the 
usual isovector pions, but the isosinglet parts play no role
in the following calculation. 
As we show below (following Ref.~\cite{ShWu2}),
the symmetric positioning of the condensate in (\ref{eq:expandsigma})
leads to the usual identification of the individual pion fields in $\pi$.
In particular, for the isovector fields,
the decomposition for each block is the usual one,
\begin{equation}
\pi = \left(\begin{array}{cc}
\pi^0/\sqrt2 & \pi^+ \\
\pi^- &-\pi^0/\sqrt2 \end{array}\right)
\,.
\end{equation}

To show this, we next need to map the operators 
$P^\pm$, $S^0$ and $\bar q_S q_V$
of Eqs.~(\ref{eq:Ppm}), (\ref{eq:S0}) and (\ref{eq:C0conn}) 
into the effective theory.
This is a standard exercise, requiring the
introduction of scalar and pseudoscalar sources into the mass matrix $M$. 
At LO in our power counting, the results are the same as in the continuum.
In particular for quark bilinears we have
\begin{eqnarray}
i\bar q T \gamma_5 q &\longrightarrow& -i\frac{f^2 B_0}2
\Str\left({\cal P}_q T [\Sigma^\dagger-\Sigma]\right)
\end{eqnarray}
and
\begin{eqnarray}
\bar q T q &\longrightarrow& -\frac{f^2 B_0}2
\Str\left({\cal P}_q T[\Sigma+\Sigma^\dagger]\right)
\,.
\end{eqnarray}
where $T$ is an arbitrary flavor matrix acting on the $4\times4$
quark subspace.
Using this result, and the expansion (\ref{eq:expandsigma}),
we find 
\begin{eqnarray}
P^\pm &\longrightarrow& -i\frac{f^2 B_0}2
\Str\left({\cal P}^{SS}\tau_{\pm}
[\Sigma_{\rm ph}^\dagger-\Sigma_{\rm ph}]\right)
\\
&=& - 2 f B_0 \pi^{\mp}_{SS} + {\cal O}(\pi^3)
\\
S^0 &\longrightarrow& \frac{f^2 B_0}2
\Str\Big({\cal P}^{SS}\big\{
c_0[\Sigma_{\rm ph}+\Sigma_{\rm ph}^\dagger] 
\nonumber\\
&& - i s_0 \tau_3 [\Sigma_{\rm ph}^\dagger-\Sigma_{\rm ph}]
\big\}\Big)
\\
&=& - s_0 2 f B_0 \pi^0_{SS} +  {\cal O}(\pi^2)
\\
-\bar q_S q_V &\longrightarrow& \frac{f^2 B_0}2
\Str\Big({\cal P}^{SV}\big\{
c_0[\Sigma+\Sigma^\dagger]
\nonumber\\ 
&&- i s_0 \tau_3 [\Sigma_{\rm ph}^\dagger-\Sigma_{\rm ph}]\big\}\Big)
\\
&=&  - s_0 2 f B_0 \pi^0_{VS} +  {\cal O}(\pi^2)
\,.
\end{eqnarray}
where $c_0=\cos\omega_0$ and $s_0=\sin\omega_0$.
${\cal P}^{SS}$ is the projector onto the sea-sea block,
and ${\cal P}^{SV}$ picks out the off-diagonal valence-sea block:
\begin{equation}
{\cal P}^{SS} =
\left(\begin{array}{ccc}
1 & 0 & 0 \\
0 & 0 & 0 \\
0 & 0 & 0 \end{array}\right)
\,,\quad
{\cal P}^{SV} =
\left(\begin{array}{ccc}
0 & 1 & 0 \\
0 & 0 & 0 \\
0 & 0 & 0 \end{array}\right)
\,.
\end{equation}
Thus, at maximal twist ($s_0=1$), 
$P^\pm$ and $S^0$ indeed couple with equal strength to the charged and
neutral pions, respectively, as required by the underlying
theory. We also see that $\pi^0_{VS}$ is the appropriate
field to use to determine the connected part of the neutral
correlator.

We can now calculate the correlators $C^\pm(x)$, $C^0(x)$ and
$C^{0,{\rm conn}}(x)$ of Eqs.~(\ref{eq:Cpm}), (\ref{eq:C0})
and (\ref{eq:C0conn}), respectively.\footnote{%
Note that we are now in a continuum theory, so the lattice
label $n$ is replaced by Euclidean position $x$
(with the correspondence $x\sim a n$).}
At this stage we specialize to maximal twist. This not only
simplifies the resulting expressions but also turns out,
as sketched in Appendix~\ref{app:nonmaxtwist}, to
give the same constraint on the LECs as one finds when
working at arbitrary twist. 
Expressed in terms of the rotated fields,
the chiral Lagrangian becomes
\begin{eqnarray}
{\cal L}_\chi &=&
\frac{f^2}4 \Str\left(
 \partial_\mu \Sigma_{\rm ph} \partial_\mu \Sigma_{\rm ph}^\dagger\right)
-
\frac{f^2}4 2 B_0 m_q 
\Str\left(\Sigma_{\rm ph}\!+\!\Sigma_{\rm ph}^\dagger\right)
\nonumber\\
&& - \hat{a}^2 W_6' \left[\Str\left(\Sigma_0\Sigma_{\rm ph}
                 +\Sigma_0^\dagger\Sigma_{\rm ph}^\dagger\right)\right]^2
\nonumber\\
&& - \hat{a}^2 W_7' \left[\Str\left(\Sigma_0\Sigma_{\rm ph}
-\Sigma_0^\dagger\Sigma_{\rm ph}^\dagger\right)\right]^2
\nonumber\\
&&
 - \hat{a}^2 W_8' \Str\left(
\Sigma_0\Sigma_{\rm ph}\Sigma_0\Sigma_{\rm ph}
+\Sigma_0^\dagger\Sigma_{\rm ph}^\dagger
 \Sigma_0^\dagger\Sigma_{\rm ph}^\dagger\right)
\,.
\label{eq:Lph}
\end{eqnarray}
Keeping only terms quadratic in the pion fields, we find that
the $W_6'$ term gives
\begin{equation}
- w'_6 f^2 \left(\pi^0_{SS}+\pi^0_{VV} + {\rm ghost}\!-\!{\rm terms}\right)^2
\,,
\label{eq:w6expand}
\end{equation}
the $W_7'$ term vanishes, and the $W_8'$ term becomes
\begin{equation}
- w'_8 f^2 \left(\frac12 [\pi^0_{SS}]^2 + \pi^0_{SV}\pi^0_{VS} +
\frac12 [\pi^0_{VV}]^2 + {\rm ghost}\!-\!{\rm terms}\right)
\,,
\end{equation}
Here we are using rescaled, dimensionless LECs
\begin{eqnarray}
w'_k = \frac{16 \hat{a}^2 W'_k}{f^4} \qquad (k=6,7,8)
\,.
\end{eqnarray}
We see that, while the $W'_8$ term contributes to the masses
of all the neutral pions, the $W'_6$ term contributes only
to the neutral particles in the diagonal blocks (and thus not to
the $\pi^0_{SV}$ mass). This is because of the double-strace
form of $W'_6$, which means that it gives ``hairpin vertices''
in the usual PQChPT parlance.

Putting this all together, we find that, at leading order,
each correlator of interest is proportional
to the propagator of the corresponding pion.
In momentum space we have (still at maximal twist)
\begin{equation}
\widetilde C^j(p) = \frac{4 f^2 B_0^2}{p^2+(m_\pi^j)^2}
\,,
\label{eq:Cjres}
\end{equation}
with $j=\pm$, $0$, and ``$0,{\rm conn}$'',
where\footnote{%
There is one subtlety in the calculation.
As can be seen from Eq.~(\ref{eq:w6expand}),
there are off-diagonal terms proportional to $w'_6$
connecting $\pi^0_{SS}$ to $\pi^0_{VV}$ and ghost terms.
These do not contribute, however, due to a cancellation
between valence and ghost contributions, as must be the
case because, for a purely sea-quark pion, we can do the
calculation solely in the unquenched WChPT, leading to
the result stated.}
\begin{eqnarray}
(m_\pi^\pm)^2 = (m_{SS}^\pm)^2 &=& 2 B_0 \mu
\label{eq:mpipm}
\\
(m_\pi^0)^2 = (m_{SS}^0)^2 &=& 2 B_0 \mu - (2 w'_6 + w'_8) f^2
\label{eq:mpi0}
\\
(m_\pi^{0,{\rm conn}})^2 = (m_{SV}^0)^2 &=& 2 B_0 \mu - w'_8 f^2
\,.
\label{eq:mpi0conn}
\end{eqnarray}
The results for $m_{\pi^\pm}$ and $m_{\pi^0}$
agree with those of Refs.~\cite{Munster,Scorzato,ShWu1},
while that for the connected neutral pion is new.
It is the latter result which
provides the key constraint, as we now explain.

\section{Mass inequality and the constraint on LECs}
\label{sec:inequalities}

We begin by rewriting the charged correlators using
$\gamma_5 G(-\mu)\gamma_5= G(\mu)^\dagger$
(which follows from $\gamma_5$-hermiticity)
\begin{eqnarray}
C^+(n) &=& 2\langle \tr(G(\mu)_{0,n}G(\mu)^\dagger_{n,0})\rangle
\\
C^-(n) &=& 2\langle \tr(G(\mu)^\dagger_{0,n}G(\mu)_{n,0})\rangle
\,.
\end{eqnarray}
These two correlators are equal by charge conjugation symmetry,
i.e. after averaging over each gauge field and its complex
conjugate. Note that both correlators are a sum over
positive definite terms, which leads us to expect that they
are larger than all other correlators
(assuming appropriate overall normalization factors).
This is the basis for the mass-inequality method. 

In the present case, we can adapt the argument given in
Ref.~\cite{BarGS}. We start by noting that, on each gauge
configuration,
\begin{eqnarray}
0 &\le& \left|\left[G(\mu)+G(\mu)^\dagger\right]_{0a,nb}\right|^2
\label{eq:ineq1}
\\
&=& \left[G(\mu)+G(\mu)^\dagger\right]_{0a,nb}
    \left[G(\mu)^\dagger+G(\mu)\right]_{nb,0a}
\,,
\end{eqnarray}
where $a$ and $b$ are color-Dirac indices.
Multiplying out, 
summing over the color-Dirac indices, 
averaging over configurations [allowed since the quark
determinant is real and positive], 
and using Eqs.~(\ref{eq:C+initial}) and (\ref{eq:C0conn_initial}), 
we arrive at the key inequality\footnote{%
The correlators $C^\pm$ are real and positive, while $C^{0,{\rm cont}}$
is {\em a priori} only known to be real but of indeterminate sign.
The PQWChPT result (\ref{eq:Cjres}) shows, however, that at long
distances $C^{0,{\rm cont}}$ is also positive. Thus we chose to
consider the sum $G(\mu)+G(\mu)^\dagger$ in Eq.~(\ref{eq:ineq1}),
so that $C^{0,{\rm conn}}$ would appear with a positive sign on
the right-hand-side of the inequality (\ref{eq:inequality}).
We note for completeness, however, that we could also have
considered the difference in $G(\mu)-G(\mu)^\dagger$ in Eq.~(\ref{eq:ineq1}),
from which one would deduce that, in general,
$C^+(n)\ge |C^{0,{\rm conn}}(n)|$.}
\begin{equation}
\frac{C^+(n) + C^-(n)}2 = C^+(n)
\ge C^{0,{\rm conn}}(n)
\,,
\label{eq:inequality}
\end{equation}
which holds for all $n$.

Now, for long distances, we can use the forms predicted by
PQWChPT, which we know from the previous section to be 
(after Fourier transforming)
\begin{eqnarray}
C^+(n) &\propto& (m_{SS}^+)^{1/2}(an)^{-3/2} e^{-m_{SS}^+ a n}
\label{eq:C+form}
\\
C^{0,{\rm conn}}(n) &\propto& (m_{SV}^0)^{1/2}(an)^{-3/2}e^{-m_{SV}^0 an}
\,,
\label{eq:C0connform}
\end{eqnarray}
with a common coefficient of proportionality.
We stress that, although $C^{0,{\rm conn}}$ is unphysical,
PQWChPT predicts that it has a single-particle exponential
fall-off at long distances.
The only way that (\ref{eq:C+form}) and (\ref{eq:C0connform})
can be consistent with the inequality
(\ref{eq:inequality}) for $n$ large enough that the
exponential damping dominates is if
\begin{equation}
m_{SS}^+ \le m_{SV}^0 
\end{equation}
or, equivalently,
\begin{equation}
m_\pi^\pm \le m_\pi^{0,{\rm conn}}
\,.
\label{eq:minequality}
\end{equation}
Combining this with the results for the masses from
PQWChPT, Eqs.~(\ref{eq:mpipm}) and (\ref{eq:mpi0conn}),
we find that
\begin{equation}
w'_8 \le 0  \ \ \Leftrightarrow \ \ W'_8 \le 0
\,.
\end{equation}
As shown in Appendix~\ref{app:nonmaxtwist}, one finds
no other constraints on the LECs if one repeats the
argument at non-maximal twist.

The inequality (\ref{eq:minequality})
can be directly tested in lattice simulations,
and present results (see, e.g., Fig.~6 of Ref.~\cite{ETM08})
clearly satisfy the inequality.

We close this section by noting a relationship between the 
mass inequality
(\ref{eq:minequality}) and the analysis of the condensate
given in Appendix~\ref{app:vafawitten}.
One of the conclusions from the appendix is that the sea-valence
SU(2) symmetry cannot be spontaneously broken for non-zero $\mu_0$.
This is consistent with the mass inequality because,
if there were a mixed sea-valence condensate, then one would
expect that fluctuations in the sea-valence direction would
diverge, and thus that $(m_{SV}^0)^2$ would pass through zero
and become negative. The mass inequality says that this
cannot happen while $m_{SS}^+$ is positive, as it is expected
to be for any non-zero $\mu_0$.

\section{Conclusions}
\label{sec:conc}

We have shown that the sign of one of the LECs induced in
Wilson ChPT by discretization errors can be determined by
combining partially quenched WChPT with mass inequalities.
The core of the argument is technically very simple, requiring
only a tree-level computation and a simple inequality.
The only connection between our argument and those given in
Refs.~\cite{DSV,ADSV} is that both require
the positivity of the determinant.

The constraint we find is that $W'_8\le 0$, independent of
the values of $W'_6$ and $W'_7$. We find no constraints on
the latter two LECs. These results are the same as
found in Ref.~\cite{gap06}, based on the failure of
a method to calculate the spectral density of the hermitian
Wilson-Dirac operator. Our constraint is also consistent
with that given in Ref.~\cite{ADSV} based on the positivity 
of the partition function in odd topological sectors
($W'_8\le 0$ if $W'_6=W'_7=0$).
It differs from that found using the convergence of
the zero-mode partition function, namely $W'_8\le W'_6+W'_7$~\cite{ADSV}.
Whether our result is stronger or weaker than this constraint depends on
the signs of $W'_6$ and $W'_7$. 

We stress that all arguments leading to constraints 
rely on the applicability
of partially quenched ChPT. In our case, we work in
the ``$p$-regime''---i.e. large volumes, with only
small perturbations around the ground state---while
Refs.~\cite{DSV, ADSV} work in the $\epsilon$-regime
in which the zero-modes must be integrated over the
entire group manifold.

Our calculation also provides a simple way of determining
$W'_8$ using the result (valid at maximal twist,
and generalized to arbitrary twist in Appendix~\ref{app:nonmaxtwist})
\begin{eqnarray}
(m_\pi^{0,{\rm conn}})^2 - (m_\pi^\pm)^2 
&=&  - w'_8 f^2 + {\cal O}(a^4, a^2 m_\pi^2)
\\
&=& - \frac{16 {\hat a}^2 W'_8}{f^2} + {\cal O}(a^4, a^2 m_\pi^2)
\label{eq:split}
\,.
\end{eqnarray}
It appears from recent simulations with twisted-mass fermions
(see, e.g., Refs.~\cite{ETM08,ETM09})
that this should give a fairly accurate determination.
The only concern is whether the LO contribution will dominate.
It would thus be interesting to extend the one-loop calculation of
Refs.~\cite{Bartmnlo,Ueda11} to the partially quenched theory.
It would also be interesting to
compare results obtained using (\ref{eq:split})
with those from other 
recently proposed methods for determining $W'_8$,
which are based on using a mixed action~\cite{Herdoiza11}
or on partially quenched pion scattering amplitudes~\cite{HansenSharpe}.

We return now to the implications for the phase structure of
unquenched twisted-mass fermions. As noted in the introduction,
this depends on the sign of the combination of LECs, $2 W'_6+W'_8$.
If this combination is negative, then one is in the Aoki-phase
scenario, which means that 
$m_\pi^\pm < m_\pi^0$ as long as $\mu_0\ne 0$
[as can been seen from Eqs.~(\ref{eq:mpipm}) and (\ref{eq:mpi0})].
The results of Refs.~\cite{Farchioni:2004us,Farchioni:2004ma,Farchioni:2004fs,%
Farchioni05,Farchioni:2005bh,Farchioni06,Boucaud:2007uk,ETM08,ETM09,%
Baron:2010bv},
however, favor the first-order scenario, with
$m_\pi^\pm > m_\pi^0$ for $\mu_0\ne 0$. This means that 
$2 W'_6+W'_8>0$, which, combined with the inequality $W'_8\le 0$,
implies in turn that $2 W'_6> |W'_8|$. There is nothing theoretically
inconsistent with this possibility, but it is somewhat
surprising given that $W'_6/W'_8 \propto 1/N_c$ for large $N_c$.

A related implication of the presence of the first-order scenario
is that quark-disconnected contributions play an important role.
It is these contributions which, despite being suppressed by $1/N_c$,
lower the neutral pion mass below that of the charged pion.
This violation of large $N_c$ counting (Zweig's rule)
is superficially analogous to the situation with the $\eta'$ in QCD.
Here, however, the effect has the opposite sign,\footnote{%
This point has been stressed recently in Ref.~\cite{CreutzECT}.}
and is of ${\cal O}(a^2)$ rather than a physical effect. 

\section{Acknowledgments}
SS thanks Gregorio Herdoiza for suggesting the calculation
of the connected neutral pion correlator, and benefited
from conversations with many of the participants at the ECT*
workshop on ``Chiral dynamics with Wilson fermions'', October 2011.
We thank Paul Damgaard, Maarten Golterman,
Gregorio Herdoiza, Karl Jansen, Kim Splittorff and Jac Verbaarschot for
comments.
This work is supported in part by the US DOE grant no.~DE-FG02-96ER40956. 
%
\appendix

\section{Form of the partially quenched condensate}
\label{app:vafawitten}

In this appendix we present the arguments for
the form of the condensate given in Eq.~(\ref{eq:Sigma0}).
The discussion is carried out in the underlying theory.

We first note that we know from a general
argument given in Ref.~\cite{BarGS}
that quark-ghost condensates vanish.
This leads to the zero entries in the rightmost column and bottom row 
(aside from the bottom-right block).

Secondly, we show that the valence-sea
$SU(2)$ symmetry is unbroken, leading to the other zeros
in (\ref{eq:Sigma0}), as well as the result that the condensate
in the valence-valence block is the same as that in the sea-sea block.
The argument is a generalization of the Vafa-Witten theorem
on the absence of flavor breaking~\cite{VafaWittenflavor}. 
A similar argument was made in Ref.~\cite{BarGS} concerning 
the absence of flavor-breaking in the valence sector alone,
but this was dependent on the fact that the valence sector
contained quarks with an exact chiral symmetry, so that
the Dirac operator has a continuum-like spectrum.
In the present case we have valence and sea Wilson fermions, with no
chiral symmetry, so the argumentation is different.
In fact, it is surprising that one can make such an argument
at all, since we know that the $SU(2)$ symmetry in
the sea-sector {\em can} be spontaneously broken---this is, after all,
what happens in the Aoki phase. The key difference here is that we
are working at non-vanishing twisted mass, which avoids the
appearance of small eigenvalues of Wilson-Dirac operator.

We show first that the sea-valence condensate
\begin{equation}
\langle \bar q_{Su} \gamma_5 q_{Vu} \rangle
\label{eq:condSV}
\end{equation}
vanishes. The notation here is that, in each $2\times2$
block, we label the two states by $u$ and $d$.
Thus $q_{Vu}$ is the valence $u$ quark.
We work in a volume $V$, at non-zero lattice spacing $a$,
and with $\mu_0$ non-zero. We turn on a source term
\begin{equation}
{\cal L}_{\rm source} = \Delta\, \bar q_{Vu} \gamma_5 q_{Su}
\,,
\end{equation}
chosen to ``push'' the condensate in a direction
such that (\ref{eq:condSV}) is non-zero.
We then take $V\to\infty$, followed by $\Delta\to 0$,
and find that (\ref{eq:condSV}) vanishes.
This implies the absence of spontaneous symmetry breaking.

Explicitly, a simple calculation yields (up to corrections 
proportional to $\Delta^3$)
\begin{widetext}
\begin{eqnarray}
\frac1V \sum_n \langle \bar q_{Su} \gamma_5 q_{Vu}(n) \rangle
&=& \frac{\Delta}{V} \left\langle \Tr\left(
\gamma_5 \frac1{D_W+m_0+i\mu_0\gamma_5} \gamma_5
\frac1{D_W+m_0+i\mu_0\gamma_5}\right)\right\rangle
\\
&=& \frac{\Delta}{V} \left\langle \Tr\left(
\frac1{Q+i\mu_0}\frac1{Q+i\mu_0} \right)\right\rangle
\\
&=&
\Delta \int d\lambda \; \rho(\lambda) \frac{1}{(\lambda+i\mu_0)^2}
\,.
\end{eqnarray}
\end{widetext}
Here the traces are over space, Dirac and color
indices, $Q=\gamma_5 (D_W+m_0)$ is the hermitian Wilson-Dirac operator,
which has (real) eigenvalues denoted by $\lambda$,
and $\rho(\lambda)$ is the density of eigenvalues per unit volume
after averaging over gauge fields.
Note that we expect $\rho(0)$ to be non-vanishing in general
(which gives rise to the Aoki-phase~\cite{Aokiphase,SS})
but the presence of $\mu_0\ne0$ shields us from the potential
singularity at $\lambda=0$. Indeed, the coefficient multiplying
$\Delta$ is finite for any non-zero $a$, since the range
of the integration over $\lambda$ is finite. Thus the
sea-valence condensate vanishes when $\Delta\to 0$.

Note that to make this argument we need the eigenvalue density
to be well-defined, and for this we need 
the integration over gauge fields to have a positive weight,
which is the case for twisted-mass fermions.

Similar arguments show that
all the condensates $\langle \bar q_{Sj} \gamma_5 q_{Vk} \rangle$
vanish, with $j$ and $k$ running independently over $u$ and $d$.
Also, by using different twisted masses for valence and sea quarks
one can show that condensates
$\langle \bar q_{Sj} \gamma_5 q_{Sk} -
\bar q_{Vj} \gamma_5 q_{Vk} \rangle$ vanish.
For the corresponding scalar condensates, e.g.
$\langle \bar q_{Sj} q_{Vk} \rangle$, one ends up with
expressions such as
\begin{equation}
\frac{\Delta}{V} \left\langle \Tr\left(\gamma_5
\frac1{Q+i\mu_0}\gamma_5\frac1{Q+i\mu_0} \right)\right\rangle
\,.
\end{equation}
Although an eigenvalue decomposition cannot be used here,
there is no reason to expect that the coefficient of $\Delta$
diverges for non-zero $a$, given the presence of $\mu_0\ne 0$.
Assuming so, we find that
{\em all} sea-valence condensates vanish.

\section{Connected pion masses at arbitrary twist angle}
\label{app:nonmaxtwist}

In this appendix we give the values of the masses \((m_\pi^{\pm})^2\),
\((m_\pi^{0})^2\) and \((m_\pi^{0,\mathrm{conn}})^2\) at arbitrary
twist angle. For a lattice theory in the large cut-off effects 
regime, if the mass twist
angle \(\omega_m\) is not an integer multiple of \(\pi/2\), then it
will differ from the twist angle in the condensate, \(\omega_0\). 
As a result, when
the leading order chiral Lagrangian is expressed in the physical basis
(in terms of \(\Sigma_{ph}\)), the form of both the mass term and the
${\cal O}(a^2)$ terms is altered by the twist. Expanding the leading order
chiral Lagrangian to \(\mathcal{O}(\pi^2)\), one fixes the relation
between \(\omega_m\) and \(\omega_0\) by demanding that the linear
term vanish. This relation is the same as in the unquenched 
case~\cite{Munster,Scorzato,ShWu1}:
\begin{equation}
\label{eq:nolinear}
2 B_0 \mu \sin(\omega_0-\omega_m) = - f^2 (2w_6' + w_8') s_0 c_0
\,.
\end{equation}
This can be used to rewrite the quadratic terms 
in ${\cal L}_\chi$ so that they depend only
\(\omega_0\) and not $\omega_m$. One may then read off the masses
\begin{eqnarray}
(m_\pi^{\pm})^2 & =& \frac{2 B_0 \mu}{s_0} 
\label{eq:mpipmtw}\\
(m_\pi^{0})^2 & =& \frac{2 B_0 \mu}{s_0} - f^2 (2w'_6+w'_8) s_0^2 \\ 
(m_\pi^{0,\mathrm{conn}})^2 & =&  \frac{2 B_0 \mu}{s_0} - f^2 w_8' s_0^2\,.
\label{eq:mpiconntw}
\end{eqnarray}
The results for the unquenched charged and neutral pions agree
with those of Refs.~\cite{Scorzato,ShWu2}.

By generalizing the arguments of Section \ref{sec:inequalities} 
one can show that the connected neutral
mass can be no smaller than the charged mass.
We sketch the generalization briefly.
The form of the charged correlator, Eq.~(\ref{eq:C+initial}),
is independent of twist.
The neutral correlator does, however, depend on twist;
the operator used to create the neutral sea-valence pion becomes
\begin{equation}
\overline q_{S} i \gamma_5 \tau_3 e^{i \gamma_5 \tau_3 \omega} q_{V}
\,.
\end{equation}
Here $\omega$ is the twist angle determined in the
simulation, either from the input masses, or using one of
the other possible definitions. It will not matter which
definition is used, since the inequality holds independent of $\omega$.
Thus the connected neutral correlator becomes
\begin{eqnarray}
C^{0,{\rm conn}}_\omega(n) &=& -\Big\langle
{\rm tr}\big(i\gamma_5 e^{i\gamma_5\omega}G(\mu)_{0,n} 
i\gamma_5 e^{i\gamma_5\omega}G(\mu)_{n,0}
\nonumber\\
&+&i\gamma_5 e^{-i\gamma_5\omega}G(\mu)^\dagger_{0,n}
i\gamma_5 e^{-i\gamma_5\omega}G(\mu)^\dagger_{n,0}\big)
\Big\rangle\,.
\end{eqnarray}
Now, using 
\begin{equation}
\left|\left[-G(\mu) i\gamma_5e^{i\gamma_5\omega_0}
+ i\gamma_5 e^{-i\gamma_5\omega_0} G(\mu)^\dagger\right]_{0a,nb}\right|^2
\ge 0
\,,
\end{equation}
and following similar steps as in the main text, one finds that
\begin{equation}
C^{0,{\rm conn}}_\omega(n) \le C^+(n)\,.
\end{equation}
We stress that this inequality holds separately at each value
of the input bare masses $m_0$ and $\mu_0$, and furthermore, 
for fixed $m_0$ and $\mu_0$, it holds for any choice of $\omega$.
When we evaluate the correlators in WChPT it is most natural to
choose $\omega=\omega_0$, for then the connected neutral correlator
couples to the sea-valence neutral pion with the same strength as
the charged correlator does to the charged pion.\footnote{%
Using other values of $\omega$ leads to a weaker inequality.}
This means that the WChPT result Eq.~(\ref{eq:Cjres}) still holds, 
except that the masses which appear
are now those of Eqs.~(\ref{eq:mpipmtw}-\ref{eq:mpiconntw}) above.

Putting this all together,
it follows that
\begin{equation}
(m_\pi^{0,\mathrm{conn}})^2 - (m_\pi^{\pm})^2 = - f^2 w_8' s_0^2 +
\mathcal{O}(a^3) \ge 0
\,.
\end{equation}
Thus, on the one hand,
the result \(W_8' \le 0\) can be demonstrated using any non-zero twist
angle, but, on the other, working at arbitrary twist does not provide an
additional constraint on the LECs.

\bibliographystyle{apsrev} 
\bibliography{ref} 

\end{document}